\documentclass[10pt, Journal]{IEEEtran}
\IEEEoverridecommandlockouts
\usepackage{cite}
\usepackage{xcolor}
\usepackage{amsmath}
\usepackage{comment}
\usepackage{amssymb}
\usepackage{amsthm}
\usepackage{amsfonts}
\usepackage{algorithm}
\usepackage{subcaption}
\usepackage{import}
\usepackage{braket}
\usepackage{algpseudocode}
 \usepackage{balance}
\usepackage{graphicx}
\usepackage{soul}

\usepackage{multirow}
\begin{document}

\title{QuBEC: Boosting Equivalence Checking for Quantum Circuits with QEC Embedding}
\author{Chao~Lu,~\IEEEmembership{Student Member,~IEEE,}
        Navnil~Choudhury,~\IEEEmembership{Student Member,~IEEE,}\\
        Utsav Banerjee,~\IEEEmembership{Member,~IEEE,}
        Abdullah~Ash~Saki,~\IEEEmembership{Member,~IEEE,}
        Kanad~Basu,~\IEEEmembership{Member,~IEEE}

\thanks{Chao Lu, Navnil Choudhury, and Kanad Basu are with the Department of Electrical and Computer Engineering, University of Texas at Dallas, Richardson, GA USA. e-mail: (cxl200053,nxc210017,kxb190012@utdallas.edu)}
\thanks{Utsav Banerjee is with Department of Electronic Systems Engineering, Indian Institute of Science, Bengaluru, Karnataka, India. e-mail:utsav@iisc.ac.in}
\thanks{Abdullah~Ash~Saki is with IBM research, New York, NY. e-mail:axs1251@psu.edu}
}


\maketitle

\begin{abstract}

Quantum computing has proven to be capable of accelerating many algorithms by performing tasks that classical computers cannot. Currently, Noisy Intermediate Scale Quantum (NISQ) machines struggle from scalability and noise issues to render a commercial quantum computer. However, the physical and software improvements of a quantum computer can efficiently control quantum gate noise. As the complexity of quantum algorithms and implementation increases, software control of quantum circuits may lead to a more intricate design. Consequently, the verification of quantum circuits becomes crucial in ensuring the correctness of the compilation, along with other processes, including quantum error correction and assertions, that can increase the fidelity of quantum circuits. In this paper, we propose a Decision Diagram-based quantum equivalence checking approach, QuBEC, that requires less latency compared to existing techniques, while accounting for circuits with quantum error correction redundancy. Our proposed methodology reduces verification time on certain benchmark circuits by up to $271.49 \times$, while the number of Decision Diagram nodes required is reduced by up to $798.31 \times$, compared to state-of-the-art strategies. The proposed QuBEC framework can contribute to the advancement of quantum computing by enabling faster and more efficient verification of quantum circuits, paving the way for the development of larger and more complex quantum algorithms.
\end{abstract}

\noindent 

\section{Introduction} \label{sec:intro}

Quantum computing is poised to bring the next tectonic shift in computing. By harnessing the unique properties of quantum mechanics such as superposition, entanglement, and interference, quantum computers can speed up a certain class of problems in fields like chemistry~\cite{chemestry}, optimization~\cite{optimization}, and machine learning~\cite{lu2022survey}, beyond the abilities of most high performance classical computers. To obtain the best performance of quantum computers for solving humanity's most demanding problems, different stacks in the computing paradigm must work in unison and in an optimal fashion. It ranges from designing novel high-level algorithms to developing powerful low-level hardware. Besides algorithmic and hardware efficiencies, equally important, is a highly proficient compilation stack, and in the realities of today's NISQ hardware, the need for a performant compilation tool-chain is highly pronounced. For instance, there is typically an instruction set misalignment between the high-level algorithm and the low-level hardware. The quantum circuit representation of a quantum algorithm usually contains a wide-range of high-level instructions (quantum \emph{gates}), whereas the underlying hardware offers a limited set of universal basis gates. IBM's quantum hardware, based on superconducting qubits, offers a basis gate-set consisting of \texttt{\{`RZ', `ID', `SX', `X', `CX'\}}~\cite{qiskit}. Consequently, high-level gates of the input circuit are translated into the basis gates of the hardware. Moreover, the hardware may have other limitations, such as limited qubit connectivity. It necessitates routing operations such as \texttt{`SWAP'} gates to bring distant qubits near-by so that multi-qubit gates can be applied on them. Furthermore, the gates in present-day quantum computers are noisy, which warrant more aggressive compilation passes such as gate cancellation, to minimize the overall number of gates. All these steps culminate into a drastically different final version of the device-compliant circuit from the original input circuit. This brings up a critical question for consideration: \emph{``How do we verify that the compiled version of a circuit is logically equivalent to the input circuit?''}

While the compiler is expected to generate the most optimized version of the output circuit, we need it to preserve the logical integrity of the input circuit. Thus, the idea of \emph{quantum verification} becomes critical during the compilation process. Quantum verification is a systematic approach to compare the input circuit and its compiled version to certify both circuits are logically equivalent. 
Since quantum circuits are difficult for classical computers to simulate, quantum verification on classical computers is challenging. On the other hand, quantum verification on quantum computers suffers from the imperfect quantum hardware, limited quantum resources, and security issues~\cite{blackboxverification, classicalverification}.
There are several recent approaches~\cite{proveit,burgholzer2020advanced} on quantum verification that address some of these challenges, and lay out a blueprint for the future. Although these are pioneering works in the right direction, they can be improved to gain better scalability.
While these methods are capable of performing the quantum verification process with more than 30 qubits and over 2000 quantum gates to be verified, the time consumption increases significantly with the addition of more qubits to the quantum circuit. Moreover, additional qubits imported in the system will require higher latency, and some strategies cannot verify highly intricate benchmark circuits in less than the 60 seconds threshold. Therefore, in this paper, we propose algorithmic optimizations to make the verification process more scalable, which not only can handle circuits with larger numbers of qubits, but also accomplishes the verification task faster. Moreover, we extend our verification algorithm to be applicable for quantum error correction (QEC) as well.

QEC is an important piece in the quantum computing puzzle, as gate operations in current devices are error-prone. QEC is needed to detect and correct such errors for reliable computation. QEC adds another round of circuit transformations on top of previously mentioned compilation steps. The application of QEC code on a circuit usually inflates the qubit counts of the circuit for the same logical function. The added qubits (\emph{ancillary} qubits) create data redundancy and aids the process of error detection. Bacon-Shor, color, heavy-hexagon and surface code
are examples of some QEC codes (QECCs)~\cite{egan2021fault, ryan2021realization, sundaresan2022matching, bonilla2021xzzx}. The addition of ancillary qubit creates a mismatch in the number of qubits between the original circuit and the error-corrected circuit, and thus, makes the verification process even more challenging. To alleviate the challenge, we propose a two-step methodology for verifying quantum circuits with QECC encoded into them. The first step employs an intelligent circuit pruning pass to extract the functioning information of the quantum circuit. 
Next, the second step applies our scalable verification flow on the pruned circuit to verify the logical equivalence. To the best of our knowledge, our work is the first attempt to verify the functionality of quantum circuits with embedded QECC. To this end, we make the following contributions in this paper:
\begin{itemize}
\item We propose the Position Match verification strategy as the Decision Diagram-based verification strategy that improves the state-of-the-art verification methodology.
\item We propose a methodology that enables the verification scheme to check the original functionality of a quantum circuit without quantum error correction.
\item We implement our equivalence checking strategy, and subsequently compare it with existing verification approaches on 19 benchmark circuits. Our proposed QuBEC is seen to provide enhancement up to $271.49 \times$ for time consumption,  and up to $271.49 \times$ in Decision-Diagram nodes consumption.
\end{itemize}

The rest of this paper is organized as follows. 
Section~\ref{sec:background} provides background on quantum verification, as well as quantum error correction code. Section~\ref{sec:related} provides the related work on the quantum verification methodologies and quantum error correction approaches. Section~\ref{sec:Methodology} explains the methodology for the quantum verification strategy, as well as the framework for the QEC circuits' function verification. Section~\ref{sec:Results} presents our experimental results. Section~\ref{sec:Conclusion} summarizes the paper with concluding remarks and future directions.
\section{Background and Motivation}\label{sec:background}

\subsection{Decision Diagram-based Quantum Circuit Verification}

\begin{figure}[b!]
\vspace{-5mm}
\centering
\begin{subfigure}[b]{0.13\textwidth}
\includegraphics[width=\columnwidth]{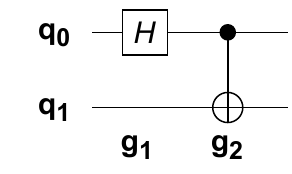}
\caption{Bell Circuit.}
\label{fig:bellcircuit}
\end{subfigure}~\begin{subfigure}[b]{0.35\textwidth}
    \centering
    \includegraphics[width =\textwidth]{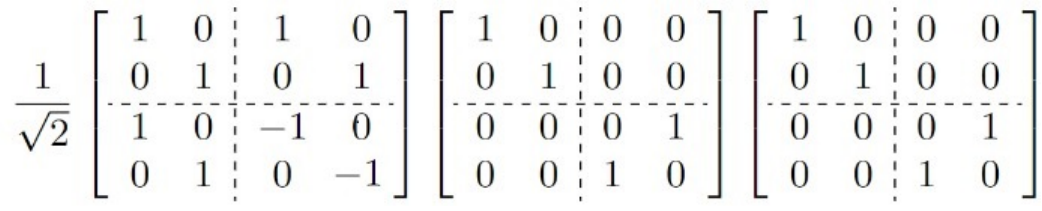}
    \caption{Matrix representation of DD.}
    \label{fig:bell matrix expression}
    \end{subfigure}
\caption{Bell State quantum circuit and its matrix representation.}
\label{bellstate}
\end{figure}

A quantum gate is a mathematical operation that acts on one or more quantum bits, noted as qubits, to perform specific operations on a quantum computer. Quantum gates can be represented as matrices. These operations can include transforming the state of the qubits, entangling them, and performing other necessary operations.

\begin{figure}[t!]
\centering
\begin{subfigure}[b]{0.16\textwidth}
    \centering
    \includegraphics[width =1.1\textwidth]{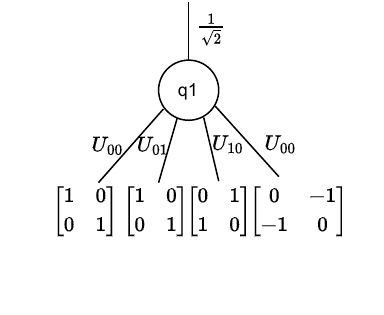}
    \caption{First level of DD.}
    \label{fig:first level}
    \end{subfigure}~\begin{subfigure}[b]{0.18\textwidth}
    \centering
    \includegraphics[width =0.55\textwidth]{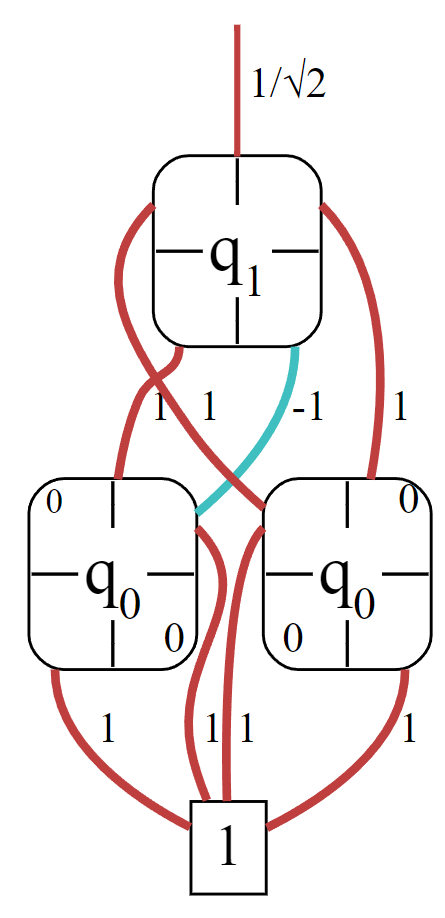}
    \caption{Complete DD.}
    \label{fig:complete dd}
    \end{subfigure}~\begin{subfigure}[b]{0.13\textwidth}
    \centering
    \includegraphics[width =0.39\textwidth]{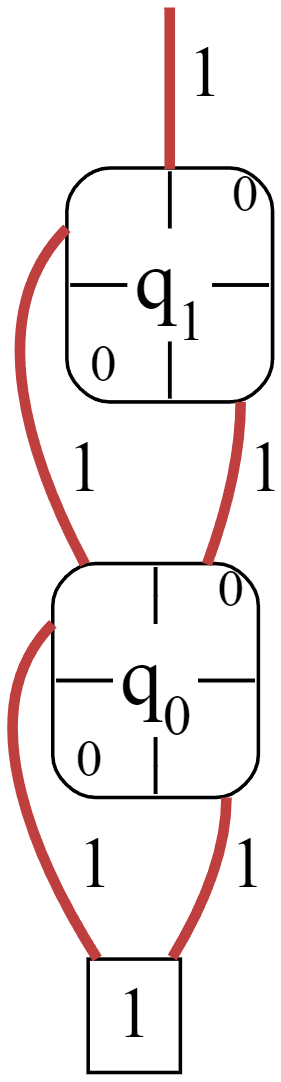}
    \caption{Identity DD.}
    \label{fig:identity dd}
    \end{subfigure}
\caption{Decision Diagram Example.}
\vspace{-5mm}
\label{fig:dd example}
\end{figure}

Decision Diagram (DD), specifically Binary DD, is used for CMOS-based classical computing verification, as it can simulate classical computing efficiently with low computing cost. Similar approach can be applied to quantum computing and simulate quantum computing circuits with relatively low cost. To represent a quantum circuit in the decision diagram form, the basic building block is a DD node, which connects multiple DD nodes to represent the quantum circuit. Fig.~\ref{fig:dd example} provides an example to illustrate the DD with corresponding DD node.

Consider the Bell State circuit shown in Fig.~\ref{fig:bellcircuit}, with its matrix representation demonstrated in Fig.~\ref{fig:bell matrix expression}. Each qubit is initialized as the ground state, represented by $\ket{0}$. The matrix expression for a quantum circuit involves a matrix form $U$ for each quantum gate. When calculating the overall matrix representation of a quantum state, each gate requires an outer product with the other qubits. Each gate has the same matrix size as the $2^n$ state, where $n$ denotes the number of qubits required by the quantum gate. For example, the matrix expression for the standard Hadamard gate is $\frac{1}{\sqrt{2}}\begin{bmatrix}1 & 1\\ 1& -1 \end{bmatrix}$. When taking the outer product of the identity matrix with $q_1$, the output is a 4 $\times$ 4 matrix, denoted as $g_1$ shown in Fig.~\ref{fig:bell matrix expression}. To obtain the final states, one can multiply all the quantum gates together in the order of ${g_n, g_{n-1}, ..., g_2, g_1}$.
To create a Bell state quantum circuit shown in Fig.~\ref{fig:bellcircuit}, two qubits ($q_0$ and $q_1$) are required. The corresponding 4 $\times$ 4 matrix is expressed in Fig.~\ref{fig:bell matrix expression}. For the overall state of the quantum circuit, a matrix multiplication operation is performed by computing $g_2$ $\times$ $g_1$. The output of the calculation is the matrix $U$ in Fig.~\ref{fig:bell matrix expression}. In the DD expression of the quantum states, the node is split into four quadrants. This operation is repeated until each quadrant only contains four elements. Following this, the DD node is generated by evaluating the elements in the smallest matrix and it is connected to the next four matrices, where each matrix forms a DD node. If the information in two matrices are the same, then the node can be merged, resulting in fewer DD nodes generated.
The first level of the DD generation is shown in Fig.~\ref{fig:first level}, and the resulting four DD matrices can be grouped into two categories of 2 $\times$ 2 unitaries: $U_{00}$ and $U_{01}$ form the first group, since they are identical, while $U_{10}$ and $U_{11}$ belong to the other group as one matrix is the negative of the other. Fig.~\ref{fig:complete dd} shows the resulting DD of the two-qubit Bell State.
The identity gate is one of the simplest quantum gates, as it does not perform any transformation on the qubit(s) it acts on, and thus, the qubit state remains unchanged. The identity gate is represented by an identity matrix,  
and its corresponding DD is shown in Fig.~\ref{fig:identity dd}.

\subsection{Quantum Gate Decomposition}

Quantum gate decomposition involves
breaking down complex quantum gates into a combination of simpler gates. This information is valuable because quantum compilers provide a library of basic gates, as discussed in Section~\ref{sec:intro}, which can be combined to achieve more complex quantum operations.

\begin{figure}[b!]
\vspace{-5mm}
\centering
\includegraphics[width=0.8\columnwidth]{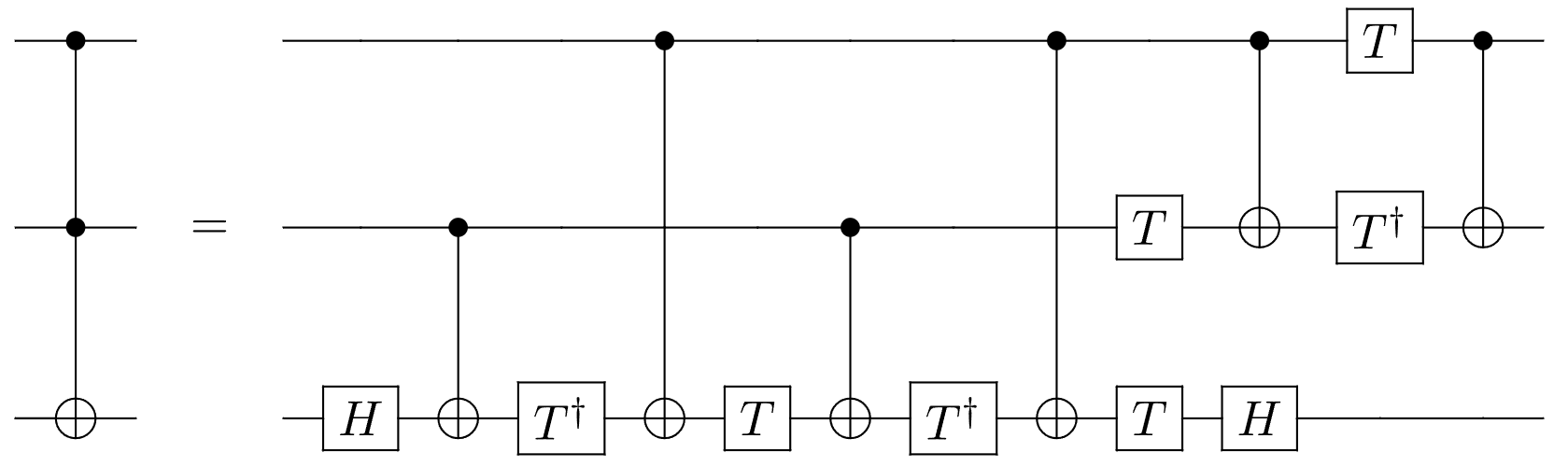}
\caption{Toffoli gate decomposition.}
\label{fig:toffolidecom}
\end{figure}

As shown in Fig.~\ref{fig:toffolidecom}, a Toffoli gate can be decomposed into a Clifford + T gate set, which constitutes another form of universal gate set for quantum computing~\cite{bocharov2015efficient}.
Quantum gate decomposition is an important tool in quantum circuit design and optimization, as it allows the efficient implementation of complex quantum algorithms using a small set of basis gates.

\subsection{DD-based Quantum Circuit Equivalence Checking} \label{sec:equivalence}

To ensure the correctness of a quantum circuit, the inverse of the quantum circuit can be appended to the original circuit. This is due to the invertible property of a quantum gate, which will generate an identity gate as each gate cancels with its inverse. However, quantum gates can cancel with each other only as long as the order of the gates is maintained. Hence, an efficient verification flow is required to verify quantum circuits so that the complex intermediate states can be avoided, by arranging the quantum circuit verification with each quantum circuit's gate order unchanged. The complexity and resource consumption of a DD-based quantum circuit simulation heavily depends on the state of qubits. Therefore, an optimized strategy for quantum circuit verification is required to address resource consumption, and improve verification efficiency on a classical computer.

To reduce resource consumption and improve verification efficiency, three alternative verification strategies and other optimization techniques were proposed in existing research~\cite{burgholzer2020advanced}.
The first strategy, the \textbf{naive approach}, involves simply alternating between two quantum circuits. However, this approach is inefficient, as it is difficult to achieve an identical DD during intermediate states in most cases. 

\begin{figure*}[tb!]
\begin{subfigure}[b]{\textwidth}
    \centering
    \includegraphics[width =\textwidth]{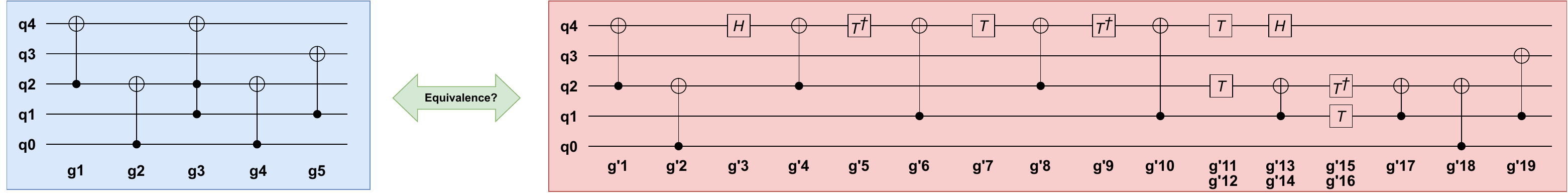}
    \caption{Example equivalence checking of two quantum circuits.}
    \label{fig:ver_sample}
\end{subfigure}
\begin{subfigure}[b]{\textwidth}
    \centering
    \includegraphics[width =\textwidth]{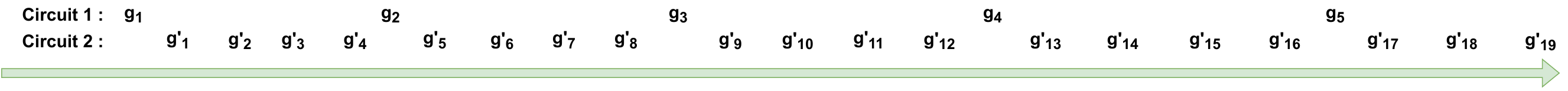}
    \caption{Quantum Proportional equivalence checking strategy.}
    \label{fig:prop}
    \vspace{3mm}
\end{subfigure}
\hfill
\begin{subfigure}[b]{\textwidth}
    \centering
    \includegraphics[width =\textwidth]{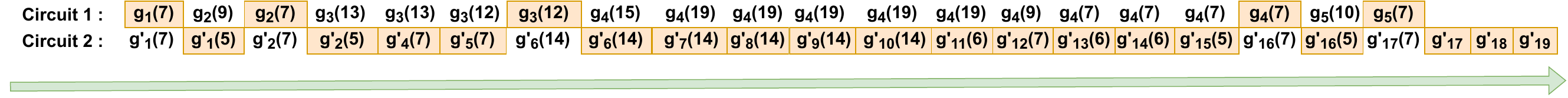}
    \caption{Quantum Lookahead equivalence checking strategy.}
    \label{fig:lookahead}
\end{subfigure}
\caption{Example showing quantum equivalence checking using existing strategies.}
\vspace{-5mm}
\end{figure*}

The second strategy, the \textbf{Proportional strategy}, is an improvement over the naive approach, where the gate counts on both sides are taken into account, and the verification proceeds proportionally. For example, if there are three gates in the first quantum circuit, and 15 gates in the second circuit, the verification will process one gate from the first circuit, and four gates from the second circuit. This strategy is considered as the most stable of the three strategies. Fig.~\ref{fig:prop} demonstrates the gate verification by using the Proportional strategy for the example circuits shown in the Fig.~\ref{fig:ver_sample}. 
If the gate type and its properties are not taken into account, it can be challenging for the strategy to achieve an identity DD in the intermediate state. This, in turn, can lead to the generation of additional DD nodes.

The third strategy is the \textbf{Lookahead strategy}, which determines the next gate to verify by calculating the computing resources required for both sides, and choosing the path that requires the minimum number of DD nodes. This method is fast for small circuits, but it tends to fall into local minima traps for larger circuits, leading to high DD node consumption. To address these issues, the open source verification library MQT/QCEC implements several additional methodologies, such as reconstructing swap gates and fusing single gates~\cite{burgholzer2020advanced}. While both Proportional and Lookahead strategies are state-of-the-art, they are difficult to scale as they cannot guarantee the optimal verification path with fewer DD nodes consumption. Fig.~\ref{fig:lookahead} demonstrates the Lookahead strategy for the verification approach. In the figure, the highlighted gates construct the selected path for the verification, and the numbers enclosed in brackets represent the node consumption of the path for the algorithm catering to the direction of verification. 
The algorithm will first determine the optimal path of the two options in the verification process, which are $g_1$ and $g'_1$. Subsequently, the algorithm will compare the node consumption of the two options and select the one which requires fewer DD nodes ($g_1$). Next, the algorithm will compare the upcoming two options, which are $g_2$ and $g'_1$, and select the path that requires the least numbers of DD nodes, which is $g'_1$. The strategy will follow a similar order until all quantum gates finish processing.

Even though the Lookahead strategy is able to find the optimal path for circuit verification, it still requires many nodes to calculate the undesired path for the strategy, which will consume a substantial amount of available computing resources. As it is shown in Fig.~\ref{fig:lookahead}, the Lookahead strategy needs to calculate the node consumption for both paths every time to determine the finest decision. However, some ``bad paths'' may subvert the computation. 
For instance, during the calculation of node consumption, the algorithm repeatedly evaluated the consumption for proceeding with $g_4$ a total of 11 times, without actually selecting $g_4$ as the chosen path. This resulted in a significant overhead in terms of DD nodes,
consequently leading to an increase in the overall computing resources needed for the task.
This path induced substantial nodes consumption and additional time overhead to finish the verification process. Additionally, the algorithm does not consider the previous quantum gates ($i.e.$, the ones already verified) to decide the strategy for the upcoming quantum gates. As a result, some important information might not be used for the verification process (contrary to our proposed Position Match approach, as explained in Section~\ref{sec:aqp}).

\subsection{Quantum Error Correction (QEC)}

Quantum error correction (QEC) is a field of study in quantum information science, and it deals with protecting quantum information from errors induced by noise and imperfect hardware~\cite{qecreview}. Unlike classical information, which can be copied and stored with insignificant errors, quantum information is very delicate and can be easily corrupted by the slightest disturbance. 


Since the pioneering work of Shor and Steane, there has been significant progress in the development of quantum error correction codes and protocols~\cite{shor1995scheme, steane2003overhead}. These developments have been driven by the increasing interest in building large-scale quantum computers, which require robust and efficient methods for error correction. Today, quantum error correction is one of the most active research areas in quantum information science and is essential for the development of practical quantum technologies.

In the context of quantum error correction, stabilizers play a key role. Stabilizers are mathematical operators that define a subspace of the Hilbert space in which a quantum system is operating. This subspace is referred to as the ``code subspace'', and it is defined in such a way that it is insensitive to certain types of errors that might affect the quantum system.

Stabilizer codes are a class of quantum error-correcting codes that are particularly useful, since they can be implemented using a relatively small number of physical qubits. The basic idea behind stabilizer codes is to encode a quantum state in such a way that the state is immune to errors that can be described by the stabilizers. Stabilizer codes are constructed using a set of stabilizer generators, which are Pauli operators that commute with each other. The set of stabilizers is chosen so that it spans a subspace that is orthogonal to the subspace spanned by the states that are affected by the stabilizers.

To understand how stabilizers work, consider a single qubit that is encoded in a stabilizer code. The stabilizer generators for this code might be the Pauli operators X and Z. The encoded state is then defined as the simultaneous eigenvector of these two operators. If a Z error occurs on the qubit, it will be detected when the stabilizer is measured, since the eigenvalue of Z will change. Similarly, an X error will be detected when X is measured. By measuring the stabilizers, it is possible to determine the type of error that has occurred and take corrective action to restore the encoded state. Stabilizer codes have many desirable properties that make them attractive for use in quantum error correction. They are easy to implement and require relatively few qubits, making them practical for use in near-term quantum computing devices. They also have high fault-tolerance thresholds, meaning that they can correct a large number of errors before the encoded state is lost. As a result, stabilizer codes are widely used in quantum computing experiments and are an important tool in the development of practical quantum technologies.
\subsection{Motivation}
Hardware verification has been a critical component in classical Integrated Circuit design flow, and similar approaches are also essential for quantum circuit design and compilation. Due to hardware restrictions and noise mitigation, the actual quantum circuit deployed in hardware may look quite different from the original design. Therefore, it is imperative to ensure their functional equivalence.

Decision Diagram (DD)-based approaches have been proposed in prior research for quantum circuit equivalence checking. However, the efficiency of the DD-based approaches varies greatly and depends heavily on the qubit phase complexity. In the best-case scenario, the efficiency can reach from exponential complexity to $\mathcal{O}(n^2)$, if the quantum circuit is simple and clean~\cite{burgholzer2020advanced}. Therefore, improving the DD is essential for verifying larger-scale quantum circuits.


Quantum Error Correction (QEC) is a critical part of the design process for quantum computing, as it helps ensure that the computations are carried out with the required level of fidelity. As part of the automation goals of quantum circuit synthesis, software is used to analyze quantum circuits and add QEC circuits to the original design in order to control quantum noise. However, current quantum circuit equivalence checking techniques are unable to verify circuits that contain redundancy with ancilla qubits and duplication, as this significantly alters the qubit and circuit function. As a result, it is crucial to find verification methods that can confirm the equivalence of quantum circuits, even when one of them contains functional redundancy. To the best of our knowledge, there are no quantum circuit verification approaches with the QEC redundancy.
\section{Related Work}\label{sec:related}


\subsection{Quantum Verification}
Various approaches have been suggested for quantum circuit verification. One such method is \textit{Prove\_it}, a quantum circuit validation algorithm that allows for the theoretical proof of the output of a quantum circuit~\cite{proveit}. However, this framework requires a significant number of hand-crafted assumptions and knowledge of quantum circuit design to validate the circuit, making it difficult to apply to new algorithms. {Moreover, the Prove-It framework does not support equivalence checking on two quantum circuits. Instead, they validate the computation of the quantum circuit in order to justify certain algorithms are correctly executed.
Another framework was proposed for designing interactive proofs that can validate the design of quantum circuits without requiring knowledge of the oracles~\cite{blackboxverification}. However, this method requires extensive knowledge of quantum mechanics and linear algebra for mathematical validation, which makes it challenging to scale up for more complex circuit designs.
DD simulation, on the other hand, is a classical simulation of quantum circuits using classical computers. It can efficiently simulate simple quantum circuits with relatively easy assumptions~\cite{hong2022tensor}. Quantum equivalence checking using decision diagrams has been proposed as a method for easier equivalence checking of quantum circuits on a classical computer~\cite{burgholzer2020advanced, burgholzer2020improved}.

\subsection{Quantum Error Correction}
To mitigate or offset the noise in quantum circuits, quantum error correction has been proposed to generate additional circuit redundancy~\cite{qecreview}. The first implementation of a quantum error correction code was developed to identify and resolve errors that occur on a single qubit within a quantum system ~\cite{shor1995scheme}. 
Dedicated attempts have been made to develop stabilizer codes for single qubit quantum error correction~\cite{gottesman1997stabilizer}. These approaches, however, neglect the consideration of the distinctive quantum hardware architecture and the limitations associated with it.
The physical architecture limitations of quantum computers necessitate the usage of numerous SWAP gates for routing. However, SWAP gates introduce considerable error rates, which are unfavorable for error correction codes. Consequently, these restrictions undermine the ability to maintain the desired low error rate in quantum systems.
To subvert this, topological QEC architectures have been proposed, which impose less stringent qubit connectivity requirements while employing a larger number of qubits~\cite{surface}.
Exisitng research has focused on finding and correcting the single-qubit-error by using Bacon-Shor, color, heavy-hexagon and surface code~\cite{egan2021fault, ryan2021realization, sundaresan2022matching, bonilla2021xzzx}.
Recently, a quantum circuit with extremely low error rate with the third generation of Sycamore quantum processor has been proposed~\cite{google2023suppressing}. 
Consequently, the redundancy provided by QEC has become an essential component in the design and implementation workflow of quantum circuits. Currently, research is being conducted to enhance the performance of QEC, for enabling its seamless automated integration with quantum circuits.
\section{Proposed QuBEC} \label{sec:Methodology}

In this paper, we propose QuBEC, a boosted quantum circuit equivalence checking scheme with QEC embedding. In this section, we initiate the discussion by examining the intricate details of our proposed Position Match strategy. This is motivated by the versatility inherent in our strategy, enabling its applicability to circuits both with and without QEC. Subsequently, we proceed to outline the implementation of our strategy, specifically taking into account the inclusion of QEC embedding.

The proposed approach stores the positions of previously verified gates, and uses a greedy method to determine the optimal path for the upcoming gates
by utilizing this information. For the remainder of this section, we will use the following terminologies:  the \emph{upcoming gate} refers to the gate that is awaiting verification, while the \emph{previous gate} is the gate that has just been verified in the last step. For example, in Fig.~\ref{fig:ver_sample}, if gates $g_1$ and  $g'_1$ are already verified, they are considered \emph{previous gates}, while gates $g_2$ and $g'_2$ are considered the \emph{upcoming gates}.
\emph{Processing} a gate refers to the operation that the program will take the matrix of the quantum gate and append to the quantum circuit. 
When two gates of two different quantum circuits are deemed equivalent, we refer to them as \emph{cancelled}. 
The following subsections explain our proposed QuBEC framework.

\subsection{Proposed Position Match Approach} \label{sec:aqp}

The proposed Position Match approach is applied to perform equivalence checking between two quantum circuits. To this end, it requires checking the positions of unverified quantum gates on both quantum circuits. To facilitate this, we introduce an active qubit pool to store the current active qubits, which are qubits that the previously verified quantum gates used. This increases the likelihood of cancellation between gates, and ensures that the quantum gates executed on inactive qubits are not processed before the ones that are already in the active pool. The size of the active qubit pool can also affect the equivalence checking flow. If the capacity is too small, complex quantum gates such as the multiple-controlled-X (MCX) gate, which contains a high number of qubits, might exceed the pool's maximum size. As a result, the MCX gate might never be checked for equivalence, or some qubits' information of the MCX gate will be lost. Conversely, if the active qubit pool is too large, several quantum gates that aren't in the same position might be processed together, reducing verification efficiency and increasing the number of intermediate nodes. 

To optimize the active qubit pool size,
a variable size approach can be adopted instead of using a fixed size pool. This approach enables the active qubit pool size to fit best for the current verification technique. We determine the pool size based on the upcoming quantum gate of the verification process. Specifically, the size of the active qubit pool will be determined by the largest quantum gates that require the most number of qubits of the verification.

\begin{figure}[!bt]
\vspace{-3mm}
\begin{algorithm}[H]
\raggedright\textbf{Input:} Two quantum circuits that is pending to be verified (circ1, circ2).\\
\raggedright\textbf{Output:} Boolean: Whether two quantum circuits are equivalent.
\caption{Position Match Strategy.}
\begin{algorithmic}[1]
\State pool = set with maximum number of(max (size(qubits required for the gate in circ1), size(qubits required for the gate in circ2)));
\State process one gate of circ1;
\State pool.add(qubits required for the gate in circ1)
\While {remaining gates on both circuits to proceed}
\State set1 = qubits required for the gate in circ1;
\State set2 = qubits required for the gate in circ2;
\State maxSize = max(length(set1), length(set2));
\If {(set1 in pool) \& not(set2 in pool)}
\State Proceed with the gate in circ1;
\ElsIf  {not(set1 in pool) \& (set2 in pool)}
\State proceed with the gate in circ2;
\Else 
\State Compare the nodes consumption with circ1, circ2 and select the best path with least nodes consumption;
\If {pool is not full}
\State pool.add(proceeded gate's corresponding qubits)
\Else 
\State pool = set(max size(qubits required for the gate in circ1), size(qubits required for the gate in circ2));
\State pool.add(proceeded gate's corresponding qubits)
\EndIf
\EndIf
\EndWhile
\While {One circuit finishes the equivalence checking}
\State Proceed the remaining gates;
\EndWhile
\If{the resultant DD is identity DD}
\State \Return True;
\Else 
\State \Return False;
\EndIf
\end{algorithmic}
\label{alg:positionmatch}
\end{algorithm}
\vspace{-0.3in}
\end{figure}

Algorithm~\ref{alg:positionmatch} presents our proposed Position Match approach to accelerate the quantum circuit equivalence checking process. The algorithm takes as input two quantum circuits, \textit{circ1} and \textit{circ2}, that need to be verified, and applies several optimization methodologies, including ``fuseSingleQubitGate", ``reconstructSWAPs", ``removeDiagonalGatesBeforeMeasure", and ``reorderOperations" from the MQT/QCEC library~\cite{burgholzer2020advanced}. Once the optimization is complete, the proposed Position Match algorithm is executed. The equivalence checking starts by processing the first gate of \textit{circ1} and adding the qubits used in the gate to the active qubit pool, with the maximum size set to the size of the input gate [Lines 1-3]. For the while loop in the algorithm, the verification has two options to proceed: the second quantum gate in \textit{circ1} or the first inverse gate in \textit{circ2} [Lines 4-7]. The algorithm checks whether the qubits required for both gates are in the active qubit pool. If one gate is in the pool and the other is not, the algorithm prioritizes the gate that is in the active qubit pool [Lines 8-9]. If both gates are not in the pool, the algorithm performs a greedy node check for both paths and selects the path that requires fewer DD nodes using the required algorithm [Lines 10-13]. After the path is selected, the active qubit pool is updated with the new qubits from the gates that were just verified [Lines 14-15]. If the maximum size of the active qubit pool is reached, the algorithm removes the previous pool and creates a new pool with a maximum size equivalent to the highest number of qubits required among the gates that are pending verification [Lines 16-19]. The new qubits of the previous gate are added to the active qubit pool. The process is repeated until all the quantum gates in one quantum circuit are processed [Line 20]. Once the checking of one quantum circuit is concluded, the algorithm will stop using the active qubit pool and process the rest of the gates in the other circuit. [Lines 22-24]. Once completed, the algorithm will compare the two circuits for equivalence by checking whether the resultant DD is an identity DD [Lines 25-29].

As mentioned in Section~\ref{sec:equivalence}, existing methods cannot address the problems of  ``gate cancellation'' and ``bad paths''. Our proposed Position Match strategy with the active qubit pool approach can address these issues with negligible cost. For example, to explain the size determination of the active qubits pool, if the quantum gates pending on the first quantum circuit contain a three-qubit gate, and the quantum gates pending on the second quantum circuit contain a single-qubit gate, then the size of the active qubit pool will be three. This means that only a maximum of three qubits will be in the active qubit pool. If the pending quantum gates contain qubits that are not in the active qubit pool, the equivalence checking will not proceed with this quantum gate until all gates that are in the active qubit pool finish computation. With this approach, the quantum gates will have a higher chance of cancelling with each other and form identity gates in the intermediate steps, while maintaining minimal calculations for the next node. This will improve the efficiency of the equivalence checking.

Now, we will explain our proposed approach using the example shown in Fig.~\ref{fig:ver_sample}. The first gate to generate the DD is $g_1$. The next step in the verification can be either $g_2$ or $g'_1$. Since the last gate that was processed, $g_1$, contained qubits $q_2$ and $q_4$, these two qubits will be added to the active qubit pool.
When the next gate is to be processed, the active qubit pool will check if all the qubits are in the pool. If the condition is true, then the gate that is in the pool will be processed first, since gates with qubits in the same pool have a higher chance of canceling with each other. Moreover, this check only requires position information, rather than the generation of decision nodes, which saves computational resources and generates fewer nodes.
\begin{figure}[!bt]
\centering
\includegraphics[width=1\columnwidth]{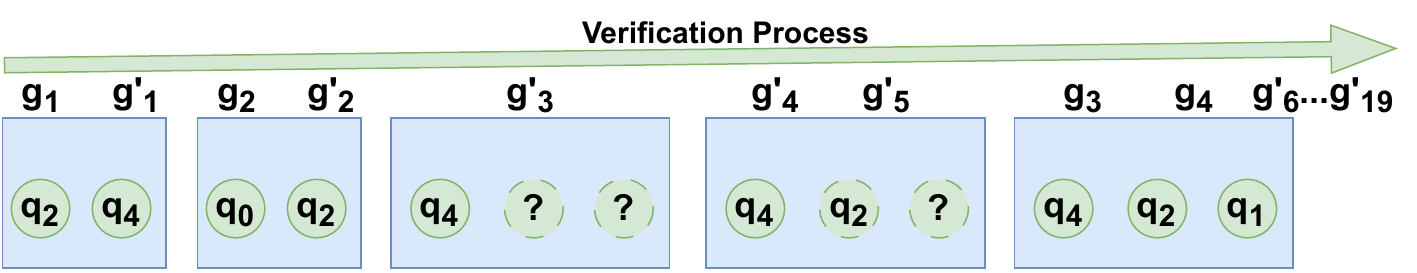}
\caption{Proposed Position Match equivalence checking strategy demonstrating the active qubit pool.}
\vspace{-0.2in}
\label{fig:aqp}
\end{figure}
\begin{figure*}[htb!]
\centering
\begin{subfigure}[b]{1\textwidth}
    \centering
    \includegraphics[width=0.96\columnwidth]{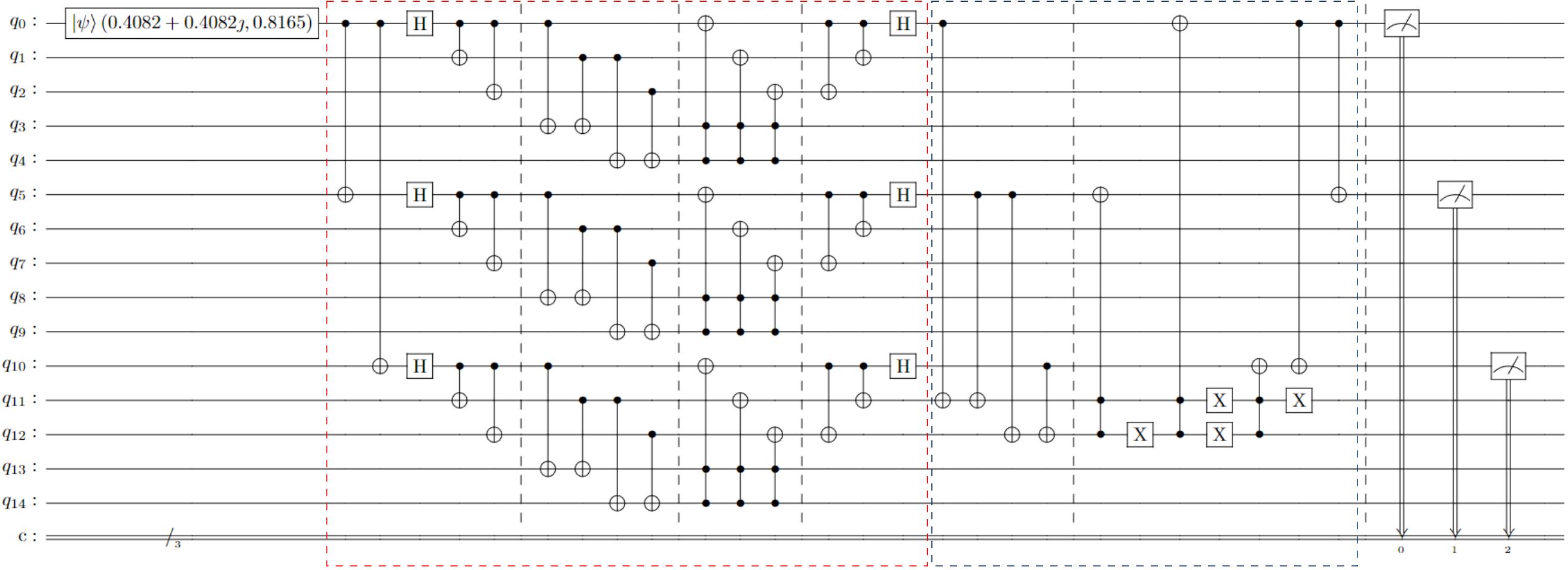}
    \caption{Quantum Circuit with QEC redundancy.}
    \label{fig:shor}
    \end{subfigure}
\begin{subfigure}[b]{0.392\textwidth}
    \centering
    \includegraphics[width =\textwidth]{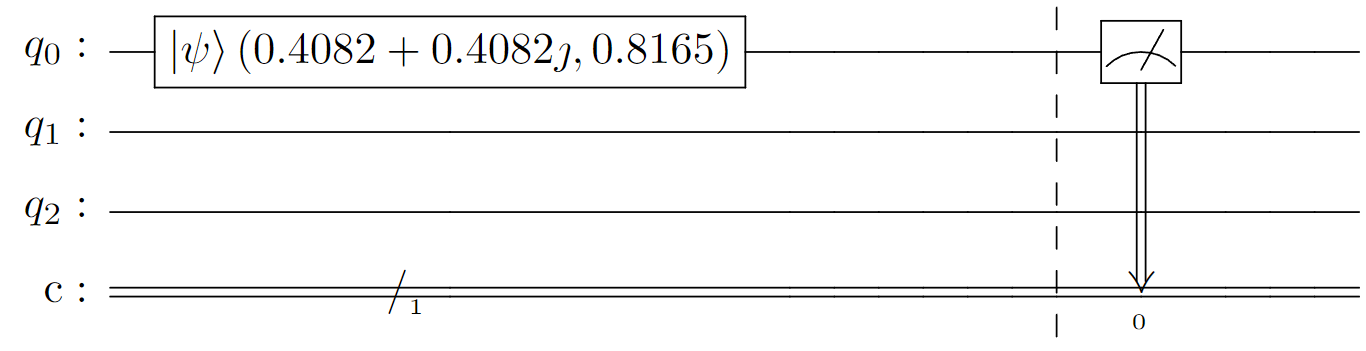}
    \caption{Original Circuit.}
    \label{fig:origin}
    \end{subfigure}
    \begin{subfigure}[b]{0.52\textwidth}
    \centering
    \includegraphics[width =\textwidth]{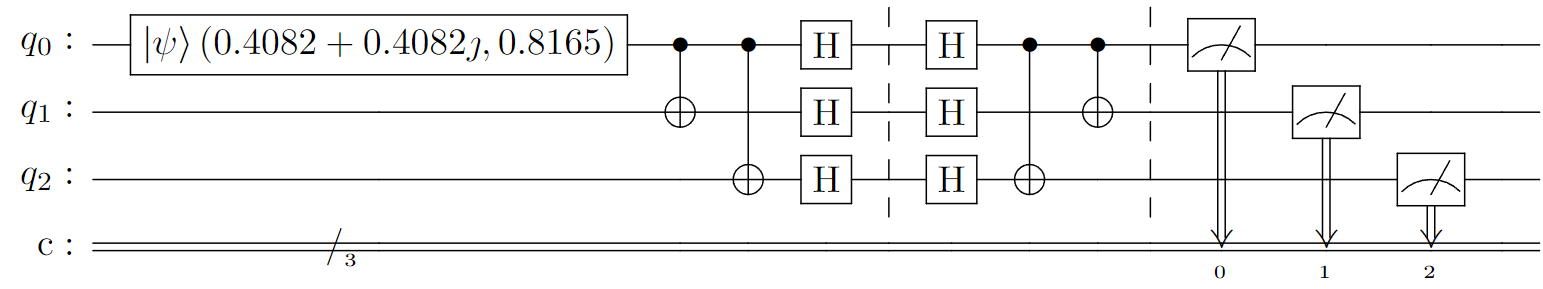}
    \caption{Pruned Quantum QEC circuit.}
    \label{fig:pruned}
    \end{subfigure}
    \caption{Proposed verification strategy for QEC-based circuits.}
    \label{fig:shor9Q}
\vspace{-5mm}
\end{figure*}

Once $g'1$ finishes computation, there is no gate in the active qubit pool. At this point, we will check both pending gates and select the direction that generates fewer nodes in the circuit. In this case, we select $g_2$ as the next gate to process because both directions generate the same number of nodes. By iterating using the same strategy, we proceed with the gates in the following sequence: $g_1, g'_1, g_2, g'_2, g'_3, g'_4, g'_5, g_3, g_4, g'_6, ..., g'_{19}$, which is the optimal verification sequence for this specific example. To determine the size of the active qubit pool, we choose the size to be the maximum number of qubits required for the pending gates. For example, if the pending gates on one side require three qubits and the other side requires two qubits, then the size of the active qubit pool will be three. Thus, the active qubit pool's size is variable and can fit best for the current verification approach, providing an efficient and optimized verification strategy.

Fig.~\ref{fig:aqp} shows the resultant quantum pool generated using the proposed verification strategy. The functional equivalence of the two quantum circuits can be determined based on the inverse property of quantum gates. Specifically, if appending the inverse of one quantum circuit to the other quantum circuit generates the initial state of the decision diagram, as shown in Fig.~\ref{fig:identity dd}, the two circuits can be considered equivalent.

\subsection{Quantum Verification with Quantum Error Correction}
Currently, there are various QEC design principles proposed for different architectures~\cite{qecreview}. However, QECC is still heavily dependent on manual design for specific quantum circuits and hardware architectures. As a result, it is challenging to automatically verify whether redundancy and additional circuitry are encoded for QEC. In this paper, we introduce a verification flow that integrates the QEC algorithm with our proposed DD-based quantum equivalence checking methodology. This allows us to verify whether the original quantum circuit is functionally identical to the final circuit, which includes the error correction code. Therefore, once the automation of the QEC is implemented, the proposed verification protocol can validate the QEC circuit without knowing its functionality.

Fig.~\ref{fig:shor} shows the original circuit for Shor's Error correction circuit~\cite{shor1995scheme}. 
Most quantum error correction code (QECC) approaches involve duplicating and measuring the original qubits, alongside the introduction of ancilla bits to provide redundancy for correcting bit-flip and phase-shift errors. The measured bits are intended to serve the same purpose as the original qubits, enabling the algorithm to determine the correct result based on the majority of the measured outcomes.
Based on the specification of the QECC, the error correction redundancy does not interfere or change the original computation output. Hence, we can assume that the QEC redundancy on the original circuit will not contain any gates to change the information of the qubits, if all quantum gates on the other qubits are pruned.
Therefore, the pruned circuit will have the same functionality as the original without any redundancy for quantum noise. For separate qubit states, the duplication method is relatively simpler, compared to entangled states. Fig.~\ref{fig:origin} demonstrates the qubit state duplication circuit for three qubits, which can be accomplished easily by inserting CX gates. Please note that this pruning technique has to be performed after the quantum circuit is reshaped by reconstructing SWAP gates, since the information of qubits can be changed by using SWAP gates, which may prune out important information and keep the redundancy in the resultant circuit.

To clarify the input requirements for the quantum circuit verification approach, information regarding the duplicated qubits in the output of the QEC code that correspond to the original circuit's qubits are needed. For example, in the case of Shor's QEC circuit, the original circuit contains only one qubit, whereas the counterpart of the error-corrected circuit contains three duplicated qubits which will be measured at the end, namely $q_{00}, q_{01}, q_{02}$, as demonstrated in Fig.~\ref{fig:shor9Q}. Therefore, the qubits in the original quantum circuit will correspond to the measured qubits in the output of the QEC circuit.

Following this, we apply our proposed verification strategy. 
In this paper, we mainly focus on the separable state error correction. A separable state is the opposite of an entangled state, where a qubit is not entangled with other qubits. However, quantum superposition is allowed for the separable state QEC code.
The code duplication, shown in Fig.~\ref{fig:origin}, can be easily verified when the original qubit is in a superposition state. This is because the qubit state can be copied by appending CX gates to the original circuit to create the duplication codes required to correct errors. Therefore, we propose to prune the QEC circuit by removing the qubits that are not measured. The resulting circuit only contains the functional and duplication codes, as shown in Fig.~\ref{fig:pruned}. Since the qubit duplication technique is fixed and requires minimal CX gates to complete, we append the code duplication to the original circuit to create a pruned circuit that is functionally identical to the original state of the circuit. The overall flow of the proposed verification framework is illustrated in Fig.~\ref{fig:qec}.

\begin{figure}[b!]
\vspace{-0.1in}
\centering
\includegraphics[width=0.96\columnwidth]{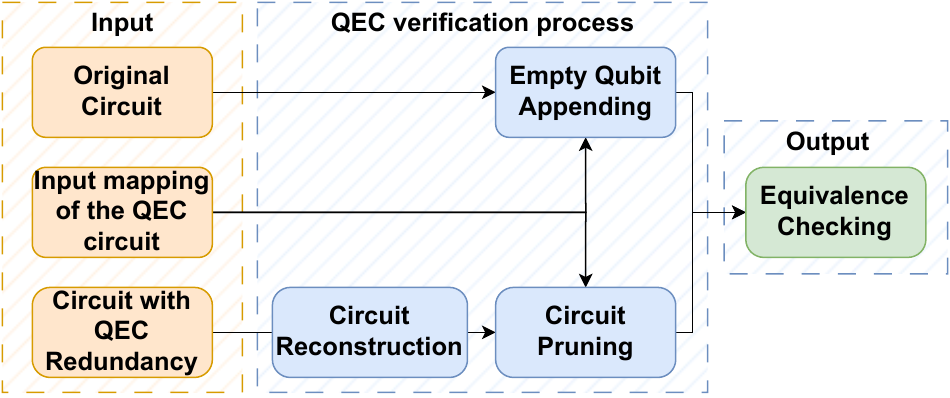}
\caption{QEC circuits equivalence checking framework.}
\label{fig:qec}
\end{figure}

As an example to illustrate the pruning approach, consider an original circuit consisting of a single qubit $q_0$ in a superposition state with a phase initialization $\ket{\psi}$. The error corrected code is shown in Fig.~\ref{fig:shor9Q}, where qubits $q_0$, $q_5$, and $q_{10}$ are subject to code duplication with a total of four CX gates. The section of the circuit marked with red corrects bit-flip errors, while the section marked in blue corrects phase-shifting errors. After computation, the qubits $q_0$, $q_5$, and $q_{10}$ are measured to obtain the error-corrected output. The duplicated qubits then vote for the majority output. To verify the equivalence of the original 15-qubit QEC circuit, the duplicated qubits and gates are pruned out, leaving only the function codes on both sides of the QEC circuit. The resulting circuit can be verified using our proposed Position Match approach, mentioned in Section~\ref{sec:aqp}. After pruning the QEC circuit, the original circuit will automatically append two empty qubits to match with the pruned 3-qubit circuits to perform the equivalence checking for the equivalence checking scheme even when two quantum circuits does not have the same number of qubits.
\section{Experimental Results} \label{sec:Results}


In this section, we demonstrate the results of our experiments for evaluating our proposed QuBEC approach. We evaluate the results furnished by our approach by conducting a comparative study against the state-of-the-art approaches, developed by~\cite{burgholzer2020improved}.

\subsection{Experimental Setup}
In this paper, our proposed strategy is based on the open source code MQT/QCEC, which is available in GitHub. We conducted the experiments on Windows 11 Enterprise 22H2 version with an Intel Xeon W-2265 CPU processor. The experiments were implemented  using g++ version 11.3.0 compiler and CMake version 3.22.1 on a VirtualBox virtual machine with Ubuntu 22.04.1 LTS, with 6 CPU cores assigned to the virtual machine.
To evaluate the performance of our proposed framework, we conducted experiments on 19 benchmark circuits utilized in existing state-of-the-art research~\cite{burgholzer2020improved}. The results obtained were then compared with those of the previous equivalence checking strategies.
\begin{table*}[tb!]
\caption{Comparison of our proposed Position Match strategy against existing equivalence checking algorithms.}
\label{tab:verification}
\scalebox{0.674}{
\begin{tabular}{|l|cccccc|cccccc|}
\hline
\multirow{2}{*}{\textbf{\begin{tabular}[c]{@{}l@{}}Benchmark\\ Circuits\end{tabular}}} & \multicolumn{6}{c|}{\textbf{Time Consumption (ms)}} & \multicolumn{6}{c|}{\textbf{Nodes Overhead}} \\ \cline{2-13} 
& \textbf{Lookahead} & \textbf{Proportional} & \textbf{Position Match} & \textbf{Best Strategy} & \begin{tabular}[c]{@{}c@{}}\textbf{Acceleration}\\ \textbf{over}\\ \textbf{Lookahead}\end{tabular} & \begin{tabular}[c]{@{}c@{}}\textbf{Acceleration}\\ \textbf{over}\\ \textbf{Proportional}\end{tabular} & \textbf{Lookahead} & \textbf{Proportional}   & \textbf{Position Match} & \textbf{Best Strategy} & \begin{tabular}[c]{@{}c@{}}\textbf{Acceleration}\\ \textbf{over}\\ \textbf{Lookahead}\end{tabular} & \begin{tabular}[c]{@{}c@{}}\textbf{Acceleration}\\ \textbf{over}\\ \textbf{Proportional}\end{tabular} \\ \hline
dk27\_225,                                                                             & 97        & 45           & 47            & Proportional  & 2.06 $\times$                                              & 0.96 $\times$                                                 & 225       & 565            & 240            & Lookahead     & 0.94 $\times$                                              & 2.35 $\times$                                                 \\
pcler8\_248,                                                                           & 109       & 38           & 54            & Proportional  & 2.02 $\times$                                              & 0.70 $\times$                                                 & 681       & 497            & 220            & \textbf{PositionMatch} & 3.10 $\times$                                              & 2.26 $\times$                                                 \\
5xp1\_194,                                                                             & 173       & 183          & 138           & \textbf{PositionMatch} & 1.25 $\times$                                              & 1.33 $\times$                                                 & 971       & 1482           & 728            & \textbf{PositionMatch} & 1.33 $\times$                                              & 2.04 $\times$                                                 \\
alu1\_198,                                                                             & 113       & 69           & 39            & \textbf{PositionMatch} & 2.90 $\times$                                              & 1.77 $\times$                                                 & 457       & 1330           & 636            & Lookahead     & 0.72 $\times$                                              & 2.09 $\times$                                                 \\
mlp4\_245,                                                                             & 485       & 393          & 314           & \textbf{PositionMatch} & 1.54 $\times$                                              & 1.25 $\times$                                                 & 2243      & 1131           & 381            & \textbf{PositionMatch} & 5.89 $\times$                                              & 2.97 $\times$                                                 \\
dk17\_224,                                                                             & 188       & 166          & 122           & \textbf{PositionMatch} & 1.54 $\times$                                              & 1.36 $\times$                                                 & 777       & 636            & 311            & \textbf{PositionMatch} & 2.50 $\times$                                              & 2.05 $\times$                                                 \\
add6\_196,                                                                             & 1873      & 1674         & 794           & \textbf{PositionMatch} & 2.36 $\times$                                              & 2.11 $\times$                                                 & 4388      & 2838           & 1434           & \textbf{PositionMatch} & 3.06 $\times$                                              & 1.98 $\times$                                                 \\
C7552\_205,                                                                            & 213       & 445          & 131           & \textbf{PositionMatch} & 1.63 $\times$                                              & 3.40 $\times$                                                 & 458       & 422            & 344            & \textbf{PositionMatch} & 1.33 $\times$                                              & 1.23 $\times$                                                 \\
cu\_219,                                                                               & 134       & 118          & 84            & \textbf{PositionMatch} & 1.60 $\times$                                              & 1.40 $\times$                                                 & 763       & 654            & 384            & \textbf{PositionMatch} & 1.99 $\times$                                              & 1.70 $\times$                                                 \\
example2\_231,                                                                         & 711       & 594          & 379           & \textbf{PositionMatch} & 1.88 $\times$                                              & 1.57 $\times$                                                 & 1550      & 982            & 528            & \textbf{PositionMatch} & 2.94 $\times$                                              & 1.86 $\times$                                                 \\
c2\_181,                                                                               & 460       & 175          & 89            & \textbf{PositionMatch} & 5.17 $\times$                                              & 1.97 $\times$                                                 & 2787      & 1036           & 357            & \textbf{PositionMatch} & 7.81 $\times$                                              & 2.90 $\times$                                                 \\
rd73\_312,                                                                             & 389       & 251          & 57            & \textbf{PositionMatch} & 6.82 $\times$                                              & 4.40 $\times$                                                 & 2283      & 8679           & 869            & \textbf{PositionMatch} & 2.63 $\times$                                              & 9.99 $\times$                                                 \\
cm150a\_210,                                                                           & 8340      & 433          & 90            & \textbf{PositionMatch} & 92.67 $\times$                                             & 4.81 $\times$                                                 & 426,300   & 8270           & 534            & \textbf{PositionMatch} & 798.31 $\times$                                            & 15.49 $\times$                                                \\
cm163a\_213,                                                                           & 189       & 140          & 74            & \textbf{PositionMatch} & 2.55 $\times$                                              & 1.89 $\times$                                                 & 816       & 1121           & 387            & \textbf{PositionMatch} & 2.11 $\times$                                              & 2.90 $\times$                                                 \\
sym9\_317,                                                                             & 217       & 828          & 54            & \textbf{PositionMatch} & 4.02 $\times$                                              & 15.33 $\times$                                                & 1355      & 43,448         & 460            & \textbf{PositionMatch} & 2.95 $\times$                                              & 94.45 $\times$                                                \\
mod5adder\_306,                                                                        & 2654      & 60,000       & 221           & \textbf{PositionMatch} & 12.01 $\times$                                             & 271.49 $\times$                                               & 95,318    & Did not Finish & 3724           & \textbf{PositionMatch} & 25.60 $\times$                                             & N/A                                                                        \\
rd84\_313,                                                                             & 23,019    & 11,967       & 235           & \textbf{PositionMatch} & 97.95 $\times$                                             & 50.92 $\times$                                                & 248,935   & 275,939        & 5147           & \textbf{PositionMatch} & 48.37 $\times$                                             & 53.61 $\times$                                                \\
cm151a\_211,                                                                           & 345       & 199          & 133           & \textbf{PositionMatch} & 2.59 $\times$                                              & 1.50 $\times$                                                 & 1882      & 2994           & 700            & \textbf{PositionMatch} & 2.69 $\times$                                              & 4.28 $\times$                                                 \\
apla\_203                                                                              & 415       & 319          & 265           & \textbf{PositionMatch} & 1.57 $\times$                                              & 1.20 $\times$                                                 & 911       & 828            & 410            & \textbf{PositionMatch} & 2.22 $\times$                                              & 2.02 $\times$                                                 \\ \hline
\end{tabular}
}
\end{table*}
\subsection{Result of Position Match Verification Strategy} \label{sec:resultposition}

\begin{figure}[b!]
\vspace{-5mm}
\centering
\includegraphics[width=1\columnwidth]{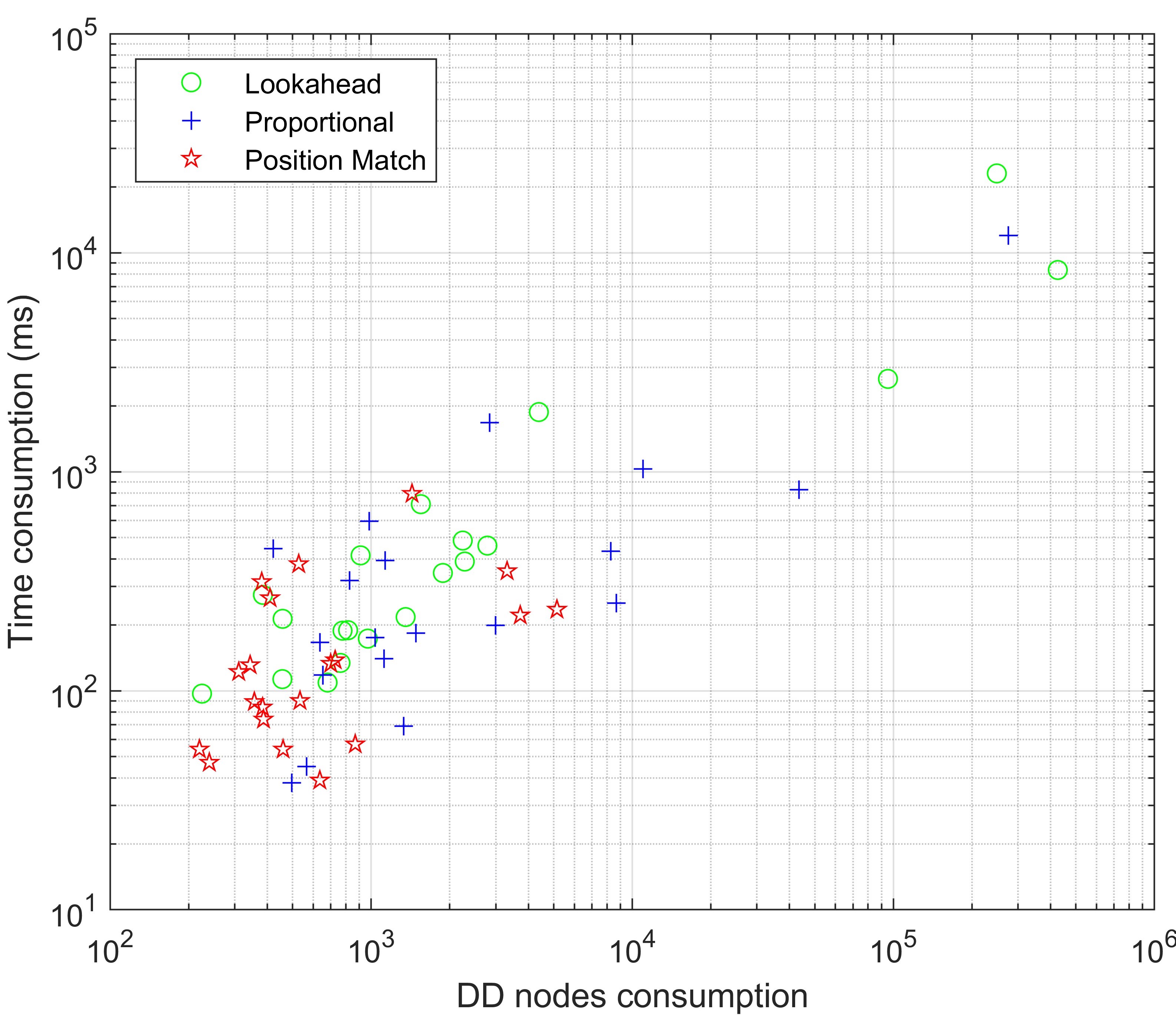}
\caption{DD nodes and time consumption of Lookahead, Proportional, and Position Match strategies.} 
\label{fig:swapinsertion}
\end{figure}

In this section, we demonstrate the results obtained by implementing our proposed strategy, Position Match on the benchmarks used in~\cite{burgholzer2020advanced}. 
Table \ref{tab:verification} presents the results for the 19 benchmark circuits considered in this study, for all three verification strategies, Proportional and Lookahead, mentioned in Section~\ref{sec:equivalence}, and the proposed Position Match approach. The best technique for each benchmark is determined based on the DD nodes consumption, and time consumption.
Column 1 consists of the benchmarks we used to evaluate our proposed Position Match strategy. Columns 2 to 7 demonstrate the time consumption. Columns 2 through 4 correspond to runtime of Lookahead, Proportional, and Position Match techniques, respectively. Column 5 indicates the best verification strategy for each benchmark used. Columns 6 and 7 demonstrate the improvement in \textit{acceleration} furnished by our Position Match strategy over Lookahead and Proportional strategies, respectively. Columns 8 to 13 describe results obtained corresponding to node overhead, and have identical outline as columns 2 to 7.  

As seen in Table \ref{tab:verification}, our proposed strategy performs significantly better in terms of time consumption compared with both Lookahead and Proportional approaches. Compared with Lookahead technique, our proposed approach furnishes up to $97.95 \times$ acceleration (for benchmark \textit{rd84\_313}) over the former, with an average of $11.85 \times$ acceleration, implying a reduction in time consumption by $1284.8ms$, on average. Meanwhile, Position Match strategy outperforms Proportional technique by furnishing an acceleration greater than $271.49 \times$ (for benchmark \textit{mod5adder\_306}), and an average acceleration of $19.59 \times$, implying an average reduction in time consumption by $583.16ms$.
In terms of node consumption,  our proposed approach provides substantial improvement over both existing techniques, reducing the node overhead by up to $798.31 \times $ and $94.45 \times $ over Lookahead and Proportional strategies respectively, as shown in Table~\ref{tab:verification}.

The significant improvement implies a tremendous reduction in resource consumption for the verification procedure. This can be seen in Fig.~\ref{fig:swapinsertion}, where DD nodes consumption and time consumption are plotted on the x and y axes respectively, for all the benchmark circuits. The figure reveals that time consumption is proportional to the DD nodes consumption. 
From Fig.~\ref{fig:swapinsertion}, it can also be observed that Lookahead and Proportional strategies require significantly more DD node and time consumption, which indicates the extensive computing resources demanded by these methods.
In contrast, the results generated by Position Match strategy are situated in the bottom left of the graph, implying low time and DD node consumption, and thus, superior performance in terms of both parameters. 
The proposed Position Match strategy is shown to be more effective, compared to state-of-the-art approaches, with significant reduction in both time and node overhead for 85\% of the benchmarks used.

The Lookahead and Proportional strategies outperform the Position Match strategy by a maximum of $29.63 \%$ in time consumption and $28.1 \%$ in Node overhead for just 2 of the 19 benchmarks. This is because the Proportional strategy has a very low initialization time to process. Hence, it is faster to perform the verification process in lightweight circuits even though the node consumption is higher. Moreover, since the Lookahead strategy always checks the DD nodes consumption of both paths encountered during verification, it might find the global minimum if the circuit is relatively small. It is important to note that the Position Match strategy still exhibits superior performance compared to existing approaches, for the majority of benchmark circuits. In fact, it surpasses them by more than $270.49 \times$ in time consumption and up to $797.31 \times$ in Nodes overhead. 
This improvement is especially evident for larger benchmarks, such as \textit{mod5adder\_306}, and \textit{rd84\_313}, which can be observed from Table~\ref{tab:verification}.

\subsection{Result of QEC Verification}
To verify that the quantum circuit with QECC operates correctly, it is necessary to ensure that the QEC does not alter the phase of the original gate. Therefore, it is safe to remove the QEC redundancy without changing the original function of the QEC. However, the original QCEC library only supports verification with the same number of qubits, and the verified quantum circuit has exactly the same function as the original circuit without any redundancy. It unrolls classical gates as a new qubit to verify, which results in incorrect output and stops the verification during initialization. To address this issue, we prune the QEC redundancy of the circuit and compare it with the original circuit that has the same function but without the QEC circuit redundancy. Fig.~\ref{fig:shor9Q} illustrates the original circuit, the transpiled quantum circuit with QEC redundancy and the pruned circuit. Once the circuit is pruned, the verification can be correctly implemented using our proposed approach.

\begin{table}[bt!]
\caption{Quantum Verification with QEC Stabilizer Code.}
\label{tab:qec_result}
\scalebox{0.88}{
\begin{tabular}{|c|c|c|c|c|}
\hline
\textbf{Quantum Circuit}            & \textbf{QEC Gates} & \begin{tabular}[c]{@{}l@{}}\textbf{Pruned}\\ \textbf{Cricuit's} \\ \textbf{Gates}\end{tabular} & \begin{tabular}[c]{@{}l@{}}\textbf{Verification}\\ \textbf{Time} \textbf{($\mu s$)}\end{tabular} & \begin{tabular}[c]{@{}l@{}}\textbf{DD} \\ \textbf{Nodes}\end{tabular} \\ \hline
Bit-Flip QEC Circuit       & 9         & 1                                                                   & 40                                                     & 3                                                   \\ \hline
Phase-Shifting QEC Circuit & 13        & 3                                                                   & 50                                                     & 3                                                   \\ \hline
Shor's QEC Circuit         & 46        & 11                                                                  & 130                                                     & 7                                                   \\ \hline
\end{tabular}
}
\vspace{-0.1in}
\end{table}

Our proposed approach is evaluated using three quantum circuits, as shown in Table~\ref{tab:qec_result}. The first column describes the circuit name, followed by the second and third columns, which present the number of gates in unpruned and pruned circuits, respectively, while the last two columns represent the verification time and the number of DD nodes used.
These three circuits with $9$, $13$ and $46$ quantum gates with QEC redundancy are pruned to $1$, $3$, and $11$ gates, respectively, to perform the verification with the original circuit without QEC redundancy. We also obtain the time consumption and DD nodes overhead for the equivalence checking process. 
Table~\ref{tab:qec_result} demonstrates that the proposed strategy can verify the quantum circuits within a short time interval of $40 \mu s$ and requiring as low as three DD nodes. This is because the scale of the pruned quantum circuit is smaller than the benchmark circuits shown in Section~\ref{sec:resultposition}. Please note that the proposed method requires the measured qubit to be the functional qubits only, while the ancilla qubits are not measured. This might not hold for some QEC strategies when the ancilla qubits requires measurement. We intend to address these corner cases in the future. Nevertheless, to the best of our knowledge, the proposed approach is the first step towards QEC-based equivalence checking for quantum circuit verification.
\section{Acknowledgement}
This research is supported by NSF grant \#2228725

\section{Conclusion}\label{sec:Conclusion}
In this paper, we proposed QuBEC, a boosted equivalence checking strategy with embedded quantum error correction code for quantum circuit verification. Our proposed Position Match strategy is based on active qubit pools, that reduces the decision diagram nodes overhead along with time consumption. When evaluated on $19$ benchmark circuits, Position Match algorithm outperforms state-of-the-art approaches for $17$ out of $19$ benchmarks, with up to $271.49 \times$ reduction in time consumption and $798.31 \times$ reduction in decision diagram nodes consumption. Furthermore, we propose a verification methodology that can verify quantum circuits with quantum error correction redundancy. In the future, we intend to eradicate the corner cases of the QEC redundancy verification.

\balance


\end{document}